# Anti-Collusion Digital Fingerprinting Codes via Partially Cover-Free Families

Mausumi Bose
Applied Statistics Unit
Indian Statistical Institute
203 B.T. Road, Kolkata 700 108, India
E-mail: mausumi.bose@gmail.com

Rahul Mukerjee
Indian Institute of Management Calcutta
Joka, Diamond Harbour Road
Kolkata 700 104, India
E-mail: rmuk1@hotmail. com).

*Abstract*: Anti-collusion digital fingerprinting codes have been of significant current interest in the context of deterring unauthorized use of multimedia content by a coalition of users. In this article, partially cover-free families of sets are considered and these are employed to obtain such codes. Compared to the existing methods of construction, our methods ensure gains in terms of accommodating more users and/or reducing the number of basis vectors.

## 1. Introduction

Digital fingerprinting is a technique for tracing consumers who use their multimedia contents for illegitimate purposes, such as redistribution. Anti-collusion codes (ACCs), which aim at deterring such unauthorized utilization by a coalition of users, have received considerable attention in the recent literature on digital fingerprinting. For an excellent account of various aspects of ACCs in general, along with illuminating theoretical results, we refer to the pioneering work of Boneh and Shaw (1998). Trappe et al. (2003) introduced an attractive class of ACCs, which are called AND-ACCs and defined as follows.

*Definition 1:* A code of $n$ binary vectors, each of length $v$, is called a $K$-resilient AND-ACC when every subset of $K$ or fewer code vectors combined element-wise under AND is distinct from the element-wise AND of every other subset of $K$ or fewer code vectors.

A code as in Definition 1 will be called a $(v, n, K)$ AND-ACC. It involves $v$ basis vectors, accommodates $n$ users, and has resilience $K$ in the sense of being capable of identifying $K$ or fewer colluders; see Trappe et al. (2003) for more details. Construction of AND-ACCs is an interesting combinatorial problem. A method, that works when $n \leq v(v-1)/\{(K+1)K\}$ and makes use of balanced incomplete block (BIB) designs, was given in Trappe et al. (2003). Kang et al. (2006) proposed another approach using group-divisible designs, while Yagi et al. (2007) used finite geometries for obtaining AND-ACCs. Recently, Li et al. (2009) developed a unified construction procedure based on cover-free families of sets and observed that this encompasses all the earlier methods. We refer to their paper for further related references.

As noted by all the aforesaid authors, for a given $K$, one prefers a $(v, n, K)$ AND-ACC with relatively large $n$ and small $v$ because this accommodates more users and avoids distribution of energy over a large number of basis vectors. From this perspective, we exploit the idea of partially cover-free families of sets to propose two techniques for obtaining these codes. Satisfyingly, our techniques yield AND-ACCs with larger $n$ and/or smaller $v$ than what the existing methods do.

## 2. Preliminaries and motivating examples

*Definition 2:* Let $\Omega$ be a universal set of $v$ elements and let H be a family consisting of $n$ subsets of $\Omega$.
(a) The incidence matrix of H is a $v \times n$ matrix with $(i, j)$th element 1 if the $i$th element of $\Omega$ belongs to the $j$th member of H, and 0 otherwise.

(b) H is called a *K-cover-free family*, or *K*-CFF(*v*, *n*), if no union of *K* or fewer members of H includes as a subset, i.e., covers, any member of H other than those involved in the union.

(c) H is called a *K-union distinct family*, or *K*-UDF(*v*, *n*), if all unions involving *K* or fewer members of H are distinct.

It is well-known (see Trappe et al., 2003) that a *K*-UDF(*v*, *n*) and a $(v, n, K)$ AND-ACC are coexistent, because the columns of the incidence matrix of the former are equivalent to the bit complements of the code vectors of the latter. Also, if H is *K*-CFF(*v*, *n*) then it is also *K*-UDF(*v*, *n*) and hence gives a $(v, n, K)$ AND-ACC. In Li et al. (2009), this fact was well recognized and employed to construct AND-ACCs.

A family of sets can, however, be *K*-UDF(*v*, *n*), and hence equivalent to a $(v, n, K)$ AND-ACC, even without being *K*-CFF(*v*, *n*). This is illustrated below.

*Example 1:* Let $\Omega = \{1, 2, 3\} \times \{0, 1, 2\}$, where $\times$ denotes Cartesian product of sets. Consider a family H of $n = 12$ subsets of $\Omega$ as given by

{10, 20, 30}, {10, 21, 32}, {10, 22, 31}, {11, 20, 32}, {11, 21, 31}, {11, 22, 30},
{12, 20, 31}, {12, 21, 30}, {12, 22, 32}, {10, 21, 31}, {11, 22, 32}, {12, 20, 30}.

Then H is not 2-cover-free – e.g., the union of {10, 20, 30} and {11, 21, 31} covers the set {10, 21, 31}. It is only partially 2-cover-free in the sense that the subfamily, consisting of the first nine sets, is 2-cover-free. Nevertheless, one can directly verify that H is 2-union distinct and hence yields a (9, 12, 2) AND-ACC, which is nonisomorphic to what one obtains using a BIB design as proposed in Trappe et al. (2003).

Example 1 holds out the promise of the existence of AND-ACCs equivalent to families which are only partially cover-free. Moreover, as we will see, partially cover-free families can also be used to construct (completely) cover-free families. These ideas lead to AND-ACCs that are better than the existing ones. The associated constructions are presented respectively in Theorems 1 and 2 of the next section. The following definition will be useful.

*Definition 3:* Let C = $\{(c_{j1}, ..., c_{jm}) : 1 \le j \le M\}$ be a code consisting of *M* code vectors each of length *m* and defined over an alphabet of size *s*. Then C is called *K-union distinct* (*K*-UD) if no two distinct subsets $J_1$ and $J_2$ of $\{1, ..., M\}$, satisfy $\{c_{ji} : j \in J_1\} = \{c_{ji} : j \in J_2\}$ for every $i$ $(1 \le i \le m)$, when each of $J_1$, $J_2$ has *K* or fewer elements.

It is not hard to see that if the minimum (Hamming) distance *d* of the code C satisfies
$$K(m - d) < m, \tag{1}$$
then C is *K*-UD. The following example, in the spirit of Example 1, shows that the converse is not true.

*Example 2:* Consider a code C over an alphabet of size $s = 3$ and consisting of 12 code vectors, each of length $m = 3$, as shown below:

{0, 0, 0}, {0, 1, 2}, {0, 2, 1}, {1, 0, 2}, {1, 1, 1}, {1, 2, 0},
{2, 0, 1}, {2, 1, 0}, {2, 2, 2}, {0, 1, 1}, {1, 2, 2}, {2, 0, 0}.

The minimum distance of C is $d = 1$, which does not meet (1) for $K = 2$. Still, it may be directly checked that C is 2-UD. The 12 code vectors here correspond to the 12 sets of the family in Example 1, a point which Theorem 1 below will clarify.

## 3. Methods of construction

Let $C = \{(c_{j1}, ..., c_{jm}) : 1 \le j \le M\}$ be a code consisting of *M* code vectors, each of length *m* and defined over an alphabet $\{\alpha_0, \alpha_1, ..., \alpha_{s-1}\}$ of size *s*. For $1 \le i \le m$ and $1 \le j \le M$, write $\tilde{c}_{ji} = l$ if $c_{ji} = \alpha_l$. Consider two families of subsets $\mathcal{F} = \{F(0), F(1),\ldots, F(s–1)\}$ and $\mathcal{G} = \{G(1),\ldots, G(u)\}$ of $\{0, 1, \ldots, q–1\}$. Let $\Omega = \{1, 2,\ldots, m\} \times \{0, 1,\ldots, q–1\}$ be a set of $mq$ elements, and let $\mathcal{H}(0)$ be a family consisting of the $M$ subsets

$$E_j = E_{j1} \cup \ldots \cup E_{jm}, \quad 1 \le j \le M \tag{2}$$

of $\Omega$, where, for each $j$,

$$E_{ji} = \{i\} \times F(\tilde{c}_{ji}), \quad 1 \le i \le m. \tag{3}$$

As (2) and (3) suggest, $\mathcal{H}(0)$ could equivalently be introduced via concatenation of codes but the present description will be more convenient for our proofs. We also define, for $1 \le i \le m$,

$$\mathcal{H}(i) = \{\{i\} \times G(1), \ldots, \{i\} \times G(u)\} \tag{4}$$

as a family consisting of $u$ subsets of $\Omega$. Then we have Theorems 1 and 2 below on the construction of AND-ACCs.

*Theorem 1:* If (i) $C$ is $K$-UD and (ii) $\mathcal{F}$ is $K$-union distinct, that is, a $K$-UDF($q$, $s$), then the family $\mathcal{H}(0)$ is a $K$-UDF($v$, $n$), and hence equivalent to a $(v, n, K)$ AND-ACC, where $v = mq$ and $n = M$.

*Proof:* If $\mathcal{H}(0)$ is not $K$-union distinct, then by (2) and (3), there exist two distinct subsets $J_1$ and $J_2$ of $\{1, \ldots, M\}$, such that each of $J_1$ and $J_2$ has $K$ or fewer elements and, for every $i$ ($1 \le i \le m$), the union of the sets $F(\tilde{c}_{ji})$, $j \in J_1$, equals the union of the sets $F(\tilde{c}_{ji})$, $j \in J_2$. Therefore, by condition (ii), $\{\tilde{c}_{ji} : j \in J_1\} = \{\tilde{c}_{ji} : j \in J_2\}$ and hence $\{c_{ji} : j \in J_1\} = \{c_{ji} : j \in J_2\}$, for $1 \le i \le m$. This violates condition (i).

*Remark 1*: Of special interest in the context of Theorem 1 is the case where $C$ is $K$-UD without satisfying (1). Since being $K$-UD is less stringent than meeting (1), one can hope that this would allow the use of $C$ with larger $M$ and hence yield AND-ACCs accommodating more users. A construction for such $C$ will be presented in Theorem 3 and the resulting $\mathcal{H}(0)$ will turn out to be only a partially cover-free family.

In Theorem 1, suppose $C$ is the 2-UD code of Example 2, so that $s = 3$ and, with $\alpha_l = l$ ($l = 0, 1, 2$), each $\tilde{c}_{ji}$ equals $c_{ji}$. Now, if one takes $q = 3$ and $F(l) = \{l\}$ ($l = 0, 1, 2$), then one obtains the 2-union distinct family of sets shown in Example 1.

*Theorem 2:* If (i) $K < m$, (ii) the minimum distance $d$ of $C$ satisfies $K(m – d) < m$, (iii) $\mathcal{F}$ is $K$-cover-free, that is, a $K$-CFF($q$, $s$), and (iv) no union of $K$ or fewer members of $\mathcal{F} \cup \mathcal{G}$ covers any member of $\mathcal{G}$ other than those involved in the union, then the family $\mathcal{H} = \mathcal{H}(0) \cup \mathcal{H}(1) \ldots \cup \mathcal{H}(m)$ is a $K$-CFF($v$, $n$), and hence yields a $(v, n, K)$ AND-ACC, where $v = mq$ and $n = M + mu$

*Proof:* From the definitions of $\mathcal{H}(0), \mathcal{H}(1), \ldots, \mathcal{H}(m)$, it is clear that $\mathcal{H}$ has $n = M + mu$ sets, all subsets of $\Omega$. Consider any set $H$ from $\mathcal{H}$ and any collection of $k$ sets $H_1, \ldots, H_k$ from $\mathcal{H}$ such that $k \le K$ and $H \notin \{H_1, \ldots, H_k\}$. It will suffice to show that the union of $H_1, \ldots, H_k$ does not cover $H$. Since $\mathcal{H} = \mathcal{H}(0) \cup \mathcal{H}(1) \ldots \cup \mathcal{H}(m)$, we can write

$$\{H_1, \ldots, H_k\} = \Gamma_0 \cup \Gamma_1 \cup \ldots \cup \Gamma_m, \tag{5}$$

where $\Gamma_0, \Gamma_1, \ldots, \Gamma_m$ are intersections of $\{H_1, \ldots, H_k\}$ with $\mathcal{H}(0), \mathcal{H}(1), \ldots, \mathcal{H}(m)$ respectively. Then $\Gamma_0, \Gamma_1, \ldots, \Gamma_m$ are disjoint, and if $k_0, k_1, \ldots, k_m$ be the numbers of sets in $\Gamma_0, \Gamma_1, \ldots, \Gamma_m$ respectively, then by (5), $k_0, k_1, \ldots, k_m$ satisfy

$$k_0 + k_1 + \ldots + k_m = k \le K. \tag{6}$$

First, let $H \in \mathcal{H}(0)$. Then by (2), $H = E_{j*1} \cup \ldots \cup E_{j*m}$, for some $j*$. Since $C$ has minimum distance $d$, By (2), (3), condition (iii), and the fact that $k_0 \leq K$, the $k_0$ sets of $\Gamma_0$ ($\subseteq \mathcal{H}(0)$) together cover at most $k_0(m-d)$ of the sets $E_{j*1}, \ldots, E_{j*m}$. Also, by (3) and (4), for each $i$ ($1 \leq i \leq m$), the union of the $k_i$ sets of $\Gamma_i$ ($\subseteq \mathcal{H}(i)$) can have a nonempty overlap with at most one, namely $E_{j*i}$, of the sets $E_{j*1}, \ldots, E_{j*m}$. Thus, by (5), writing $\delta_i = 1$ if $k_i > 0$ and $\delta_i = 1$ if $k_i > 0$, the union of $H_1, \ldots, H_k$ covers at most $k_0(m-d) + \delta_1 + \ldots + \delta_m$ ($= \psi$, say) of the disjoint sets $E_{j*1}, \ldots, E_{j*m}$. Now, by (6),

$$\psi \leq k_0(m-d) + k_1 + \ldots + k_m \leq k_0(m-d-1) + K. \qquad (7)$$

If $k_0 > 0$, then by (6) and condition (ii),

$$k_0(m-d-1) + K < k_0(mK^{-1} - 1) + K \leq K(mK^{-1} - 1) + K = m,$$

while if $k_0 = 0$, then by condition (i), $k_0(m-d-1) + K = K < m$, so that by (7), $\psi < m$. As a result, the union of $H_1, \ldots, H_k$ fails to cover at least one of the sets $E_{j*1}, \ldots, E_{j*m}$, i.e., this union does not cover $H$.

We next turn to the case $H \in \mathcal{H}(i)$ for some $i$ ($1 \leq i \leq m$), say $H \in \mathcal{H}(1)$. Then by (4), $H$ has empty overlap with every set in $\Gamma_2 \cup \ldots \cup \Gamma_m$. On the other hand, the union of the $k_0 + k_1$ sets of $\Gamma_0 \cup \Gamma_1$ cannot cover $H$, in view of (2)-(4), condition (iv) and the fact that $k_0 + k_1 \leq K$ (vide (6)). This proves the result.

*Remark 2:* In Theorem 2, we do not need $\mathcal{F} \cup \mathcal{G}$ to be $K$-cover-free as per Definition 2(b). While it has a $K$-cover-free subfamily $\mathcal{F}$, a union of $K$ or fewer members of $\mathcal{F} \cup \mathcal{G}$ can cover a member of $\mathcal{F}$ not involved in the union, and still the theorem will remain valid. Thus the family $\mathcal{F} \cup \mathcal{G}$, which is only partially cover-free, leads to a (completely) cover-free family $\mathcal{H}$ and, as seen in Remark 4 below, this entails gains.

*Remark 3:* Binary codes of constant weight and orthogonal arrays of index unity can help in finding $\mathcal{F}$, $\mathcal{G}$ and $C$, as stipulated in Theorem 2. This is briefly indicated below.

(a) Let $B(q, N, d, w)$ denote a binary code with minimum distance $d$ and $N$ code vectors each of length $q$ and weight $w$. Define $A(q, d, w)$ as the largest possible $N$ in a $B(q, N, d, w)$, given $q$, $d$ and $w$. Tables showing exact values of or lower bounds on $A(q, d, w)$ appear in Brouwer et al. (1990) and Smith et al. (2006), and can be used in checking the availability of a $B(q, N, d, w)$. Suppose there exist a $B(q, s, d_1, w_1)$ ($= B_1$, say) and a $B(q, u, d_2, w_2)$ ($= B_2$, say) such that

$$K(w_1 - \tfrac{1}{2}d_1) < w_1, \quad K(w_2 - \tfrac{1}{2}d_2) < w_2 \quad \text{and} \quad w_2 > Kw_1. \qquad (8)$$

Write the code vectors of $B_1$ and $B_2$ as columns to form $q \times s$ and $q \times u$ matrices and take these as the incidence matrices of $\mathcal{F}$ and $\mathcal{G}$ respectively. Then the inequalities in (8) ensure conditions (iii) and (iv) of Theorem 2.

(b) An orthogonal array $OA(s^t, m, s, t)$ of index unity is an $s^t \times m$ array, with entries from a set of $s$ symbols, such that each ordered $t$-tuple of symbols occurs exactly once as a row in every $s^t \times t$ subarray. Barring the trivial case $t = 1$, if $K(t-1) < m$ and the rows of the array are taken as the code vectors of $C$, then following MacWilliams and Sloane (1977, p. 328), conditions (i) and (ii) of Theorem 2 are met.

*Remark 4:* Under condition (ii) and with $q = s$, $F(l) = \{l\}$ ($0 \leq l \leq s-1$), Li et al. (2009) showed $\mathcal{H}(0)$ to be a $K$-CFF($ms$, $M$). Theorems 1 and 2 strengthen their findings in two directions. First, in these theorems, $q$ is potentially smaller than $m$, thereby leading to a reduction in the number of basis vectors. Second, Theo-rem 1 requires $C$ to be only $K$-UD thus allowing a larger $M$ than in Li et al. (2009), while, with the same $M$ as in their paper, Theorem 2 yields a cover-free family $\mathcal{H}$ which is larger than $\mathcal{H}(0)$, so that the AND-ACCs arising from these theorems can accommodate more users. Later, in Examples 3-6, these points are illustrated. Incidentally, the idea in Theorem 2 of augmenting $\mathcal{H}(1),\ldots,\mathcal{H}(m)$ to $\mathcal{H}(0)$ is reminiscent of an adding-column technique in the construction of orthogonal arrays (see Wang and Wu, 1991), and this is new in the context of AND-ACCs.

With reference to Theorem 1, we now proceed to construct codes that are union distinct even without meeting (1). A method, which is shown to work for $K = 2$ and can potentially be extended to general $K$, is presented. Let $s$ be a prime or prime power and let $\alpha_0, \alpha_1, \ldots, \alpha_{s-1}$ be the elements of the finite field GF($s$), $\alpha_1 = 1$ being the multiplicative identity element. Suppose $3 \leq m \leq s$ and define the following row vectors of order $m$ over GF($s$):

$$\rho(0) = (1,\ldots,1), \quad \rho(i) = (\alpha_0^i, \alpha_1^i, \ldots, \alpha_{m-1}^i) \ (i = 1, 2, \ldots). \tag{9}$$

For $t \geq 2$, let $R$ be a $t \times m$ matrix with rows $\rho(i)$, $0 \leq i \leq t-1$, and $R_0$ be a $(t-1) \times m$ matrix with rows $\rho(i)$, $0 \leq i \leq t-2$. Define $U$ as the $s^t \times m$ array with rows $\xi^T R$, $\xi \in S(t)$, and $V$ as the $s^{t-1} \times m$ array with rows $\rho(t) + \mu^T R_0$, $\mu \in S(t-1)$, where $S(i)$ is the set of the $s^i$ column vectors of order $i$ over GF($s$). It is well-known (see e.g., Dey and Mukerjee, 1999, p. 37) that $U$ is an OA($s^t, m, s, t$) of index unity and hence, following Remark 3(b), yields a code satisfying (1) if $K(t-1) < m$. Theorem 3 below shows that for $K = 2$ and odd $s$, under the same condition on $t$ and $m$, the larger array

$$W = \begin{bmatrix} U \\ V \end{bmatrix}, \tag{10}$$

with $s^t + s^{t-1}$ rows and $m$ columns, represents a 2-UD code.

*Theorem 3:* If $s$ is an odd prime or prime power, $t \geq 2$ and $2(t-1) < m$, then the $s^t + s^{t-1}$ rows of $W$, interpreted as code vectors, yield a 2-UD code of size $s^t + s^{t-1}$.

Theorem 3 is proved in the appendix. With $s = m = 3$, $t = 2$ and $\alpha_l = l$ ($l = 0, 1, 2$), it yields the 2-UD code of Example 2. A more appealing application appears in Example 3 below. While a 2-UD code arising from Theorem 3 may not satisfy (1) (cf. Example 2), it has more code vectors than the one given by $U$ alone and hence yields an AND-ACC accommodating more users. We also note that if in Theorem 1, $\mathcal{F}$ is taken as a 2-cover-free family and $C$ is obtained via Theorem 3, then the resulting $\mathcal{H}(0)$ is only partially cover-free in the sense that the subfamily of $\mathcal{H}(0)$, associated with the rows of $U$, is 2-cover-free.

## 4. Examples and concluding remarks

*Example 3:* Let $K = 2$. With $s = 83$, $t = 2$ and $m = 3$, Theorem 3 yields a 2-UD code $C$ of size $M = 6972$. The code vectors of $C$ are of length $m = 3$ over an alphabet of size $s = 83$. As seen above (10), a subfamily of these code vectors, namely those corresponding to the rows of $U$, form an orthogonal array OA($83^2$, 3, 83, 2).

Next, from Table I-B in Brouwer et al. (1990), $A(20, 6, 5) \geq 84$. Thus there exists a constant weight binary code $B(20, 83, 6, 5)$ which, as in Remark 3(a), leads to a family of sets $\mathcal{F}$ that is 2-CFF(20, 83) and hence 2-UDF(20, 83).

With $C$ and $\mathcal{F}$ chosen as above, Theorem 1 yields a (60, 6972, 2) AND-ACC.

*Example 4:* Let $K = 3$. Following Remark 3(b), take $C$ as the code represented by an OA($7^3$, 7, 7, 3). Then $M = 343$, $s = m = 7$, $d = 5$, and conditions (i) and (ii) of Theorem 2 are met. Conditions (iii) and (iv) on $\mathcal{F}$ and $\mathcal{G}$ are also met if, with $q = 7$ and $u = 2$, we take $F(l) = \{l\}$ ($l = 0, 1, \ldots, 6$) and $G(1) = \{0, 1, 2, 3\}$, $G(2) = \{0, 4, 5, 6\}$. With $C$, $\mathcal{F}$ and $\mathcal{G}$ chosen as above, Theorem 2 yields a (49, 357, 3) AND-ACC.

*Example 5:* Let $K = 3$. Following Remark 3(b), take $C$ as the code given by an OA($31^3$, 7, 31, 3). Then $M = 29791$, $s = 31$, $m = 7$, $d = 5$, and conditions (i) and (ii) of Theorem 2 are met.

Next, from Tables I-B and I-H in Brouwer et al. (1990), $A(21, 6, 4) = 31$ and $A(21, 18, 13) = 1$. Thus there exist constant weight binary codes $B_1 = B(21, 31, 6, 4)$ and $B_2 = B(21, 1, 18, 13)$ satisfying (8). Hence, with $q = 21$ and $u = 1$, following Remark 3(a), one gets $\mathcal{F}$ and $\mathcal{G}$ meeting conditions (iii) and (iv) of Theorem 2.

With $C$, $\mathcal{F}$ and $\mathcal{G}$ chosen as above, Theorem 2 yields a (147, 29798, 3) AND-ACC.

*Example 6:* Let $K = 4$. Along the lines of Example 4, take $C$ as the code represented by an OA($9^3$, 9, 9, 3). Also, with $q = 9$ and $u = 2$, let $F(l) = \{l\}$ ($l = 0, 1, \ldots, 8$) and $G(1) = \{0, 1, 2, 3, 4\}$, $G(2) = \{0, 5, 6, 7, 8\}$. Then Theorem 2 yields a (81, 747, 4) AND-ACC.

*Remark 5:* The AND-ACCs in Examples 3-6 cannot be obtained by the existing methods. For instance, in each of these, $n$ exceeds $v(v-1)/\{(K+1)K\}$ and, hence the use of BIB designs, as in Trappe et al. (2003), cannot produce them. Similarly, in the respective setup of these examples, the construction in Li et al. (2009), based on OA($83^2$, 3, 83, 2), OA($7^3$, 7, 7, 3), OA($31^3$, 7, 31, 3) and OA($9^3$, 9, 9, 3), yields AND-ACCs with $(v, n, K) =$ (249, 6889, 2), (49, 343, 3), (217, 29791, 3) and (81, 729, 4); see Remark 4. Our methods lead to larger $n$ and smaller $v$ in Examples 3 and 5, and same $v$ but larger $n$ in Examples 4 and 6.

*Remark 6:* It will be of interest to extend Theorem 3 to general $K$. Initial studies suggest that for $s$ an odd prime or prime power, $K \geq t \geq 2$ and $K(t-1) < m$, the rows of $W$ in (10) should give a $K$-UD code. However, proving this is difficult. For instance, with general $K$, counterparts of the arguments in the appendix branch into too many cases and get messy. It is hoped that the present endeavor will generate further interest in this and related problems.

**Appendix: Proof of Theorem 3**

The number of coincidences between any two rows $(x_1, \ldots, x_m)$ and $(y_1, \ldots, y_m)$ of $W$ is defined as the cardinality of the set $\{i : x_i = y_i, 1 \leq i \leq m\}$, i.e., it is $m$ minus the Hamming distance between the two rows.

*Lemma 1:* The number of coincidences between any two rows of $W$ cannot exceed (a) $t-1$, if these are distinct rows of $U$, (b) $t-2$, if these are distinct rows of $V$, and (c) $t$, if one of these is a row of $U$ and the other is a row of $V$.

*Proof:* While (a) is well-known [Dey and Mukerjee, 1999, p. 36], (b) and (c) have similar proofs. For instance, if (c) does not hold, then by the definitions of $U$ and $V$, noting that the rows of $R_0$ are also rows of $R$, a subvector of $\rho(t)$, of order $t+1$, is in the row space of the corresponding $t \times (t+1)$ submatrix of $R$. This is impossible because by (9), every square submatrix, of order $t+1$, of

$$\tilde{R} = \begin{bmatrix} R \\ \rho(t) \end{bmatrix}$$

is nonsingular as $\alpha_0, \alpha_1, \ldots, \alpha_{m-1}$ are distinct.

*Proof of Theorem 3:* Since $2(t-1) < m$, i.e., $m \geq 2t-1$, it will suffice to prove the result for $m = 2t-1$.

Then $m \geq t+1$ as $t \geq 2$, and by Lemma 1, all rows of $W$ are distinct. Hence it is clear that no two distinct collections of rows of $W$, say $\{(x_{j1},...,x_{jm}): j \in J_1\}$ and $\{(x_{j1},...,x_{jm}): j \in J_2\}$, each with two or fewer rows, can have $\{x_{ji}: j \in J_1\} = \{x_{ji}: j \in J_2\}$ for every $i$ ($1 \leq i \leq m$), whenever any of the two collections has only one row or they have one row in common.

It remains to consider two collections of the form $(a,b)$ and $(x,y)$, where $a = (a_1,...,a_m)$, $b = (b_1,...,b_m)$, $x = (x_1,...,x_m)$ and $y = (y_1,...,y_m)$ are four distinct rows of $W$. If the sets $\{a_i, b_i\}$ and $\{x_i, y_i\}$ are identical for every $i$ ($1 \leq i \leq m$), then, for each $i$, either (i) $a_i = x_i$, $b_i = y_i$, or (ii) $a_i \neq x_i$, $a_i = y_i$, $b_i = x_i$. Without loss of generality, suppose (i) holds for $1 \leq i \leq m_1$, and (ii) holds for $m_1 + 1 \leq i \leq m_1 + m_2$, where $m_1 + m_2 = m = 2t - 1$. Since $m_1 \leq t$ and $m_2 \leq t$ by (i), (ii) and Lemma 1, the pair $(m_1, m_2)$ equals either $(t, t-1)$ or $(t-1, t)$.

First let $(m_1, m_2) = (t, t-1)$. Partition $R$ and $R_0$ as $R = [R_1 \; R_2]$ and $R_0 = [R_{01} \; R_{02}]$, where $R_1$ and $R_{01}$ consist of their first $t$ columns, while $R_2$ and $R_{02}$ consist of their last $t-1$ columns. Similarly, partition the row vectors $\rho(t)$, $a$, $b$, $x$ and $y$ as $\rho(t) = (\rho(t,1) \; \rho(t,2))$, $a = (a(1) \; a(2))$, $b = (b(1) \; b(2))$, $x = (x(1) \; x(2))$ and $y = (y(1) \; y(2))$, where $\rho(t,1)$, $a(1)$ etc. consist of the first $t$ elements of these vectors, while $\rho(t,2)$, $a(2)$ etc. consist of their last $t-1$ elements. Then by (i) and (ii),

$$a(1) = x(1), \quad b(1) = y(1), \quad a(2) = y(2), \quad b(2) = x(2). \tag{A.1}$$

Since $m_1 = t$, by (i) and Lemma 1, one of $a$ and $x$, say $a$, is a row of $U$ and the other, say $x$, is a row of $V$. Similarly, one of $b$ and $y$ is a row of $U$ and the other is a row of $V$. But if $b$ is a row of $V$ and $y$ is a row of $U$, then both $b$ and $x$ are rows of $V$ and, by the last equation in (A.1), they have $t-1$ coincidences, which contradicts Lemma 1(b). Thus we need to consider only the situation where $a$ and $b$ are rows of $U$, and $x$ and $y$ are rows of $V$. Then by the definitions of $U$ and $V$, there exist $\xi_1, \xi_2 \in S(t)$ and $\mu_1, \mu_2 \in S(t-1)$ such that

$$a = \xi_1^T R, \quad b = \xi_2^T R, \quad x = \rho(t) + \mu_1^T R_0, \quad y = \rho(t) + \mu_2^T R_0,$$

where

$$\xi_1 \neq \xi_2, \quad \mu_1 \neq \mu_2, \tag{A.2}$$

as $a$, $b$, $x$, $y$ are distinct. Recalling the partitioned forms of $a$, $b$, $R$ etc., the equations in (A.1) can now be expressed as

$$\xi_1^T R_1 = \rho(t,1) + \mu_1^T R_{01}, \quad \xi_2^T R_1 = \rho(t,1) + \mu_2^T R_{01},$$
$$\xi_1^T R_2 = \rho(t,2) + \mu_2^T R_{02}, \quad \xi_2^T R_2 = \rho(t,2) + \mu_1^T R_{02}. \tag{A.3}$$

As the rows of $R_0$ are also rows of $R$, we have $R_0 = QR$, for some $(t-1) \times t$ matrix $Q$. Thus

$$R_{01} = QR_1, \quad R_{02} = QR_2. \tag{A.4}$$

By (A.4), the first two equations in (A.3) yield $(\xi_1 - \xi_2)^T R_1 = (\mu_1 - \mu_2)^T QR_1$. But by (9), $R_1$, being is a $t \times t$ submatrix of $R$, is nonsingular. Therefore, $(\xi_1 - \xi_2)^T = (\mu_1 - \mu_2)^T Q$, and hence using (A.4),

$$(\xi_1 - \xi_2)^T R_2 = (\mu_1 - \mu_2)^T R_{02}. \tag{A.5}$$

On the other hand, the last two equations in (A.3) yield

$$(\xi_1 - \xi_2)^T R_2 = (\mu_2 - \mu_1)^T R_{02}. \tag{A.6}$$

Since $s$ is odd, by (A.5) and (A.6), $(\mu_1 - \mu_2)^T R_{02}$ equals the null vector. But $R_{02}$ is a $(t-1) \times (t-1)$ sub-matrix of $R_0$ and is nonsingular, by (9). Hence $\mu_1 = \mu_2$, which contradicts (A.2).

In a similar manner, a contradiction is reached when $(m_1, m_2) = (t-1, t)$. Thus the result follows.

**Acknowledgement**: This work was supported by a grant from the Indian Institute of Management Calcutta.